\documentclass[twocolumn,prb,showpacs,floatfix]{revtex4}

\usepackage{graphicx}
\usepackage{rotating}
\usepackage{amsmath}
\usepackage{amsfonts}
\usepackage{amssymb}
\usepackage{enumerate}
\usepackage{longtable}
\usepackage{dcolumn}
\usepackage{bm}

\begin{document}

\title{The fate of orbitons coupled to phonons}

\author{K.P. Schmidt$^1$}
\email{kaiphillip.schmidt@epfl.ch}
\homepage{http://marie.epfl.ch/~kpschmid}
\author{M. Gr\"{u}ninger$^2$}
\author{G.S. Uhrig$^3$}
\affiliation{$^1$Institute of Theoretical Physics, \'{E}cole Polytechnique
 F\'{e}d\'{e}rale de Lausanne, CH-1015 Lausanne, Switzerland}
\affiliation{$^2$ 2. Physikalisches Institut A, RWTH Aachen University, D-52056
 Aachen, Germany}
\affiliation{$^3$ Lehrstuhl f\"{u}r Theoretische Physik I,
Universit\"{a}t Dortmund, 44221 Dortmund, Germany}
\date{\rm\today}

\begin{abstract}
The key feature of an orbital wave or orbiton is a significant dispersion,
which arises from exchange interactions between orbitals on distinct sites.
We study the effect of a  coupling between orbitons and phonons in one
dimension using continuous unitary transformations (CUTs).
Already for intermediate values of the coupling, the orbiton band width
is strongly reduced and the spectral density is dominated by an orbiton-phonon
continuum. However, we find sharp features within the continuum and an
orbiton-phonon anti-bound state above. Both show a significant dispersion and
should be observable experimentally.
\end{abstract}



\maketitle

\section{Introduction}
In correlated electron systems, orbital degeneracy is discussed as an
interesting source for very rich physics
\cite{tokur00,khoms05,khali05}.
The orbital degeneracy may be lifted by a coupling to phonons
\cite{jahn37} and/or by superexchange processes \cite{kugel73}.
The latter give rise to an intimate connection
between spin and orbital degrees of freedom.
Spin-orbital models predict interesting ground states such as orbital order or
quantum-disordered orbital-liquid states, and one may expect novel elementary
excitations: dispersive low-energy orbital waves
termed orbitons \cite{ishih05}.
Experimentally, the observation of orbitons has been claimed on the basis of
Raman data of the orbitally ordered compounds LaMnO$_3$ \cite{saito01} and
RVO$_3$ (R=La, Nd, Y) \cite{miyas05,sugai06a}.
In LaMnO$_3$, the bosonic orbitons are treated similar to magnons in a
long-range spin-ordered state, whereas the orbital excitations of RVO$_3$ are
discussed in terms of one-dimensional (1D) fermions equivalent to the spinons
of a 1D Heisenberg chain \cite{saito01,miyas05,ishih05}.

The role attributed to phonons varies largely.
For the case of LaMnO$_3$, it has been argued that the orbiton-phonon coupling
constant $g$ is small, only giving rise to a small shift
of the orbiton band \cite{saito01,ishih05}.
The dynamical screening of orbitons by phonons and the mixed orbiton-phonon
character of the true eigen modes
has been studied by van den Brink \cite{brink01} using self-consistent
second order perturbation theory (SOPT) in the orbiton-phonon coupling $g$.
He interpreted the Raman peaks observed at about 160 meV in
LaMnO$_3$ \cite{saito01} in terms of phonon satellites of the orbiton band,
assuming weak coupling $g$. In contrast, Allen and Perebeinos \cite{allen99}
argued that the coupling $g$ is so strong that
the eigen modes can be described in terms of local crystal-field excitations
(`vibrons').
Then, the spectral density consists of a series of phonon sidebands
(Franck-Condon effect) with a center frequency $> 1$ eV \cite{allen99}.
Experimentally, the orbiton interpretation of the Raman features observed at
about 160 meV in LaMnO$_3$ has been strongly questioned based on the comparison
with infrared data \cite{gruni02c,rucka05b}, which clearly indicate that
these features should be interpreted in terms of multiphonons.

The `non-local' collective character of the orbital excitations is relevant
if the energy scale of the superexchange is larger than the coupling to the
phonons. At present, a systematically controlled quantitative description of
the gradual transition from well-defined dispersive orbitons at $g$=0 to
predominantly  `local' crystal-field excitations for strong coupling
is still  lacking. The SOPT treatment is valid at small $g$. Qualitatively, it
shows (i) a polaronic reduction of the orbiton band width \cite{brink01}
due to the dressing of the orbiton by a phonon cloud yielding a
heavy quasi-particle, and (ii) a transfer of spectral weight from the
orbiton band to phonon side-bands,
i.e., to a broad and featureless orbiton-phonon continuum (see below).
This suggests that signatures of collective behavior are rapidly washed out
with increasing $g$.

We perform a well-controlled calculation of the spectral line shape of the
orbitons at larger values of $g$ which elucidates the {\em collective}
character of the excitations.
Our study is based on continuous unitary transformations (CUTs)
\cite{glaze93,wegne94,knett00a} realized in a self-similar manner
\cite{mielk97a} in real space \cite{reisc04}.
For clarity and simplicity, we restrict
ourselves to the minimal model in one spatial dimension (1D).
We find that well-defined, dispersive features appear in the
orbiton-plus-one-phonon continuum for intermediate values of $g$.
Additionally, a sharp orbiton-phonon anti-bound state (ABS) is formed above
the continuum. Both phenomena should allow to observe dispersive signatures of
orbitons in experiment.

\section{Model}
We study the following Hamiltonian in one dimension
\begin{eqnarray}
\nonumber
 H_0 &=& \sum_i \left(\omega_{\rm orb}^{\rm ex} + \omega_{\rm orb}^{\rm JT}
 \right) c^\dagger_i c^{\phantom{\dagger}}_i
 - \frac{J}{4}\sum_i \left(c^\dagger_{i+1}c^{\phantom{\dagger}}_i
+ c^\dagger_{i-1}c^{\phantom \dagger}_i\right)
 \\
 &+& \omega_{\rm ph}\sum_i b^\dagger_i b^{\phantom{\dagger}}_i
    + 2g \sum_i  c^\dagger_i c^{\phantom{\dagger}}_i\left( b^\dagger_i +
    b^{\phantom{\dagger}}_i\right)
\label{eq:hamilton}
\end{eqnarray}
where $c$ and $b$ are bosonic operators that represent orbitons and phonons,
respectively, $\omega_{\rm orb}^{\rm ex}$ and
$\omega_{\rm orb}^{\rm JT}=4g^2/\omega_{\rm ph}$ denote the
contributions to the local orbiton energy from superexchange and
from a static Jahn-Teller deformation \cite{brink01},
$J$ the superexchange coupling constant,
$\omega_{\rm ph}$ the phonon energy, and $g$ the coupling between
orbitons and phonons.
The Jahn-Teller energy $\omega_{\rm orb}^{\rm JT}=4g^2/\omega_{\rm ph}$
results from the static distortion of the local environment of
the transition-metal ion with orbital degeneracy.
This distortion lifts the degeneracy such that the orbitals are split into
a low-lying orbital termed ground state orbital and a higher-lying one.
The orbital excitation consists in lifting an electron from the
ground state orbital into the higher-lying one. This is the effect of
the creation operator $c^\dagger$.
The superexchange $J$ acts as a hopping amplitude of the orbiton,
giving rise to a dispersion.

The Hamiltonian (\ref{eq:hamilton}) is the bosonic version of the
well-investigated 1D Holstein model of coupled electrons and phonons
yielding polarons, see e.g.\ Refs.\ \onlinecite{hohen03,fehsk06}.
At $T=0$, Eq.\ (\ref{eq:hamilton})
can be studied with a single orbiton so that its statistics does not matter
and the fermionic and the bosonic model are equivalent.

The dynamic orbiton-phonon interaction (last term in Eq.~(\ref{eq:hamilton}))
corresponds to the creation or annihilation of a local distortive phonon
if an orbiton is present. This is qualitatively the most important term
linking the orbiton and the phonon degrees of freedom
because it represents an interaction of one orbiton with one phonon.
Of course, a hybridization term linear in both the operators of the
orbiton and of the phonon will also be present.
The hybridization makes orbital effects visible in the phonon channel
and vice versa. This fact is important for assessing the
possibility to observe phonons or orbitons by certain experimental probes.
But such a hybridization term does not change the character
of the excitations qualitatively.
One may imagine that the bilinear term has been transformed away beforehand
by a Bogoliubov diagonalization.

Compared to the complex Hamiltonian studied by van den Brink for LaMnO$_3$
\cite{brink01}, our model is stripped to the minimum by allowing only
for one local (optical) phonon, by being one-dimensional, and
by omitting the bilinear hybridization term between orbitons and
phonons discussed above. The locality of the
phonon ensures that all dispersive effects result from the orbital channel.
We expect that our results will be generic for higher dimensions also because
we do not focus on particular one-dimensional aspects. For example we consider
a single orbital excitation, not a dense liquid of excitations prone to display
specific one-dimensional physics such as Luttinger-liquid behavior.
For the same reason, we do not need to consider the
formation of `biorbitons', the equivalent of the bipolarons studied in the
context of the Holstein model.
Hence we are convinced that our model contains the generic features,
while it is clear and simple enough to allow for
the controlled computation of line shapes.

In the literature on the Holstein model, the crossover
from local polarons to large polarons is treated mainly
by numerical techniques \cite{hohen03,fehsk06}.  The focus has been
on $\omega_{\rm ph} \leq W$ where $W$ is the bare electron (here: orbiton)
band width (in 1D $W$=$J$), whereas $\omega_{\rm ph} \geq W$ is reasonable
for the description of orbitons in transition-metal oxides
\footnote{Oxygen bond-stretching vibrations with typically $\omega_{\rm ph}\!
\approx \! 80$ meV constitute the most relevant phonon mode.}.
Then, the continua of different number of phonons do not overlap in energy,
allowing for the formation of bound states below the scattering states
or of anti-bound states above them.
Our calculation elucidates the regime $\omega_{\rm ph} \geq W$ and has the
merit to provide an explicit interpretation of the features found. Thereby,
the understanding of the nature of the excitations is enhanced.

\begin{figure}[t!]
    \begin{center}
     \includegraphics[width=0.9\columnwidth,clip=]{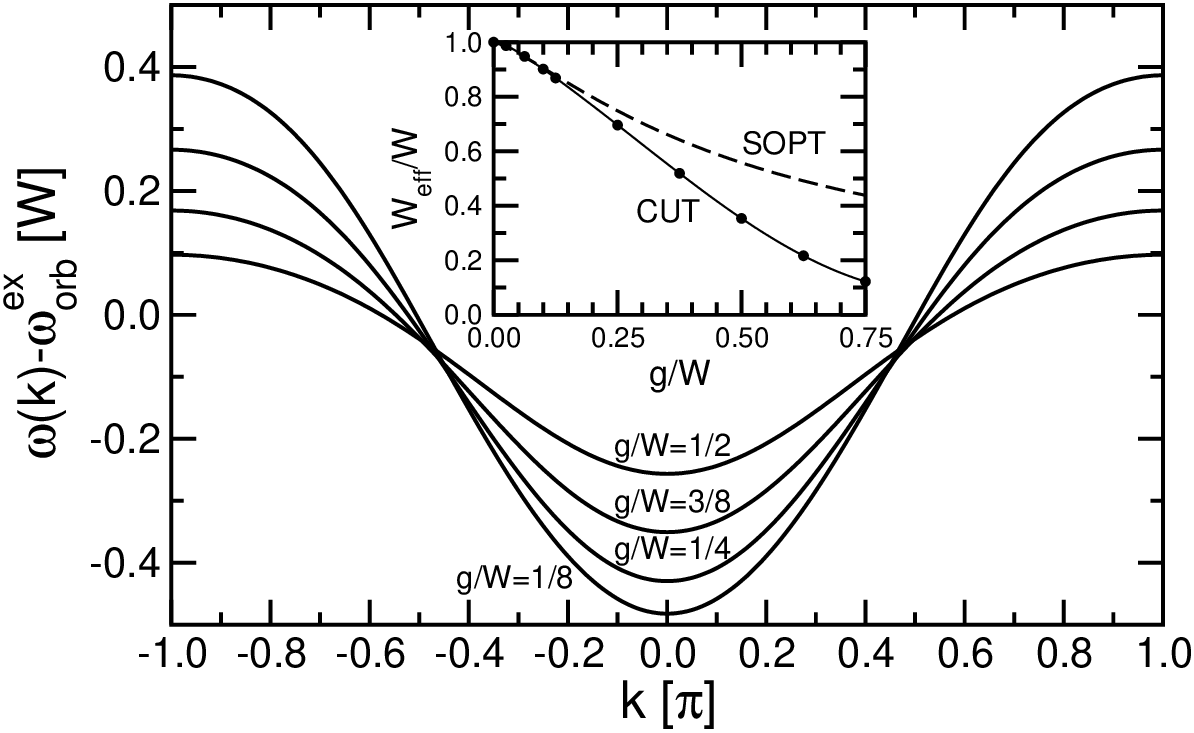}
    \end{center}
    \caption{Dispersion $\omega(k)$ of the dressed orbitons for
      $\omega_{\rm ph} = W$ using the truncation $({^2V},e\!=\!9)$.
      {\it Inset}: Effective band width $W_{\rm eff}$
      as obtained by CUT and by SOPT. For small values of $g$
      one has $\delta W_{\rm eff} \propto -g^2$ for $W< \omega_{\rm ph}$ and
      $\delta W_{\rm eff} \propto -g^{4/3}$ for $W= \omega_{\rm ph}$ in 1D.
      }
    \label{orbi-disp}
\end{figure}
\begin{figure}[t!]
    \begin{center}
     \includegraphics[width=0.9\columnwidth,clip=]{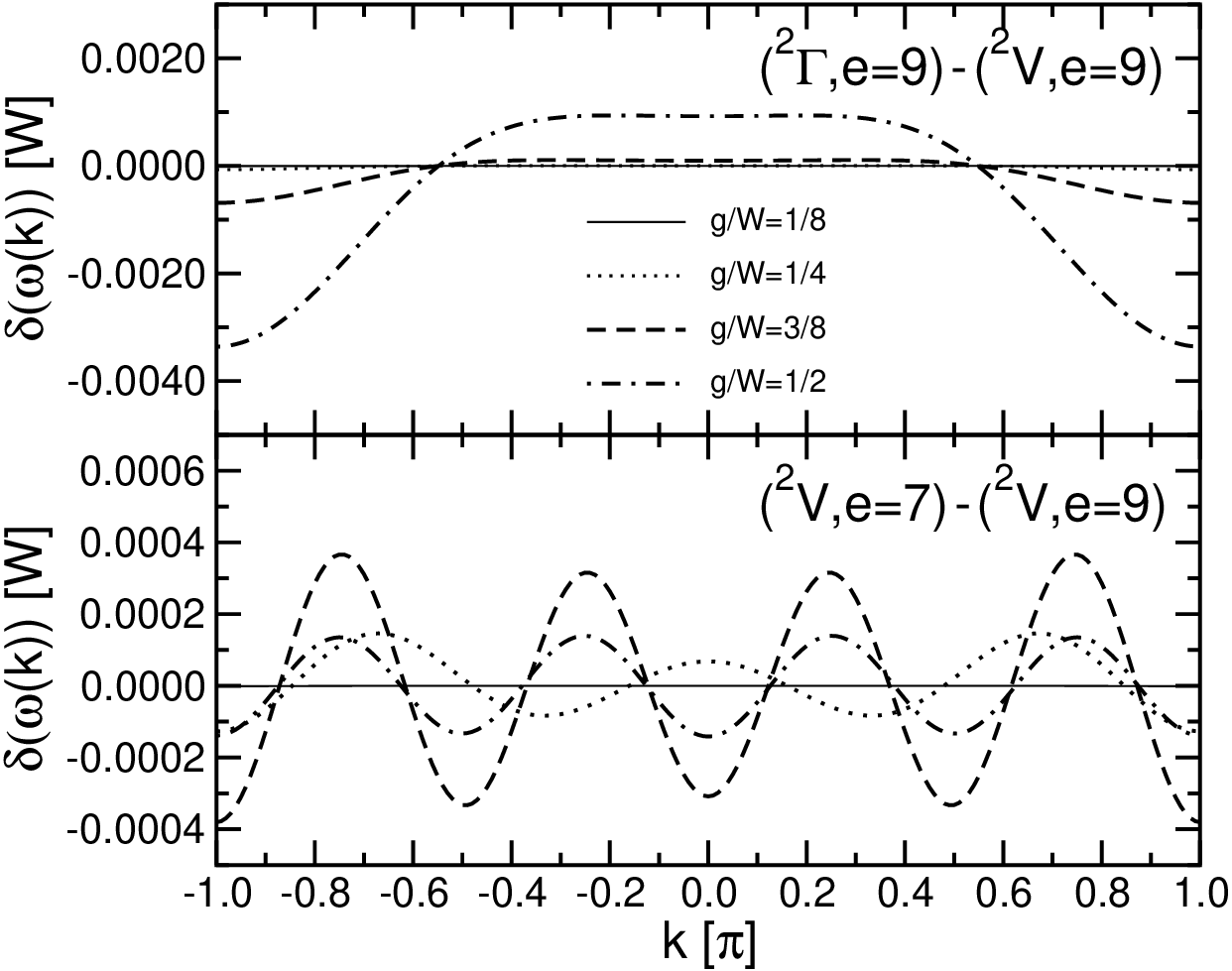}
    \end{center}
    \caption{Effect of the different truncation schemes on the orbiton
      dispersion
    $\omega(k)$ for $\omega_{\rm ph} = W$.
    Upper panel: the difference
    $\delta\omega(k) = \omega(^2\Gamma,e\!=\!9) \, - \, \omega(^2V,e\!=\!9)$
    at constant extension $e\!=\!9$.
    Lower panel:
    $\delta\omega(k) = \omega(^2V,e\!=\!7) \, - \, \omega(^2V,e\!=\!9)$
    at constant structure $S= {^2V}$.
    }
    \label{orbi-disp_truncations}
\end{figure}

\section{Method}
We choose the CUT approach \cite{glaze93,wegne94,knett00a,reisc04} because it
is an analytical approach which provides an effective model that can be
understood more easily. The CUT is defined by 
\begin{equation}
\label{eq:flow}
 \partial_l H (l) = [\eta (l),H(l)]
\end{equation} 
to transform $H(l=0)=H_0$ from its initial form to an effective Hamiltonian
$H(\infty)$ at $l=\infty$. 
Here $l$ is a \emph{continuous} running variable which 
parametrizes the unitary transformation $U(l)$ yielding 
$H(l)= U(l)^\dagger H U(l)$. The infinitesimal 
anti-Hermitian generator of the transformation is denoted by $\eta(l)$. 
The properties of the 
effective Hamiltonian depend  on the choice of the generator. 

In this work a quasi-particle-conserving CUT is used which maps the Hamiltonian
 $H_0$ to an effective Hamiltonian $H(\infty)$
which \emph{conserves} the number of 
quasi-particles\cite{knett00a,reisc04}. This is suggested from the physics of 
our model. The free orbital wave will be dressed by the interaction with the 
phononic degrees of freedom. The elementary excitations of the interacting 
model are orbiton excitations with a cloud of phonons renormalizing the 
properties of the bare orbital waves. 

The following choice for the matrix elements
of the infinitesimal generator $\eta$  
\begin{equation}
\label{eq:generator}
 \eta_{i,j} (l)={\rm sgn}\left(q_i-q_j\right)H_{i,j}(l)
\end{equation}
in an eigen basis of $H_{\rm ph}=\omega_{\rm ph}\sum_i b^\dagger_i 
b^{\phantom{\dagger}}_i$ is appropriate. 
The $q_i$ are the eigen  values of $H_{\rm ph}$\cite{knett00a}. 
Obviously, the flow (\ref{eq:flow}) stops if either $q_i=q_j$ or
$H_{i,j}=0$ for pairs of $i$ and $j$. Given that there is convergence
for $l\to\infty$, i.e.\ $H(\infty)$ exists, 
this tells us that $H(\infty)$ is block diagonal, 
i.e.\ it conserves the phonon number.

The convergence is not as easy to see generally. But assuming an almost
diagonal, non-degenerate Hamiltonian with
$q_i\leq q_j \Leftrightarrow H_{i,i} \leq H_{j,j}$
the leading order in the non-diagonal matrix elements fulfills
\begin{subequations}
\begin{eqnarray}
\partial_l H_{i,j} &=&
 - {\rm sgn}\left(q_i-q_j\right)(H_{i,i}-H_{j,j})H_{i,j}(l) \\
&=& -|H_{i,i}-H_{j,j}| H_{i,j}(l)
\end{eqnarray} 
\end{subequations}
implying convergence according to 
\begin{equation}
H_{i,j} \propto \exp(-|H_{i,i}-H_{j,j}|l)\ .
\end{equation}
The general derivations of convergence are given in Refs.\ 
\onlinecite{mielk98,knett00a}.

\begin{figure}[tb]
    \begin{center}
     \includegraphics[width=0.9\columnwidth,clip=]{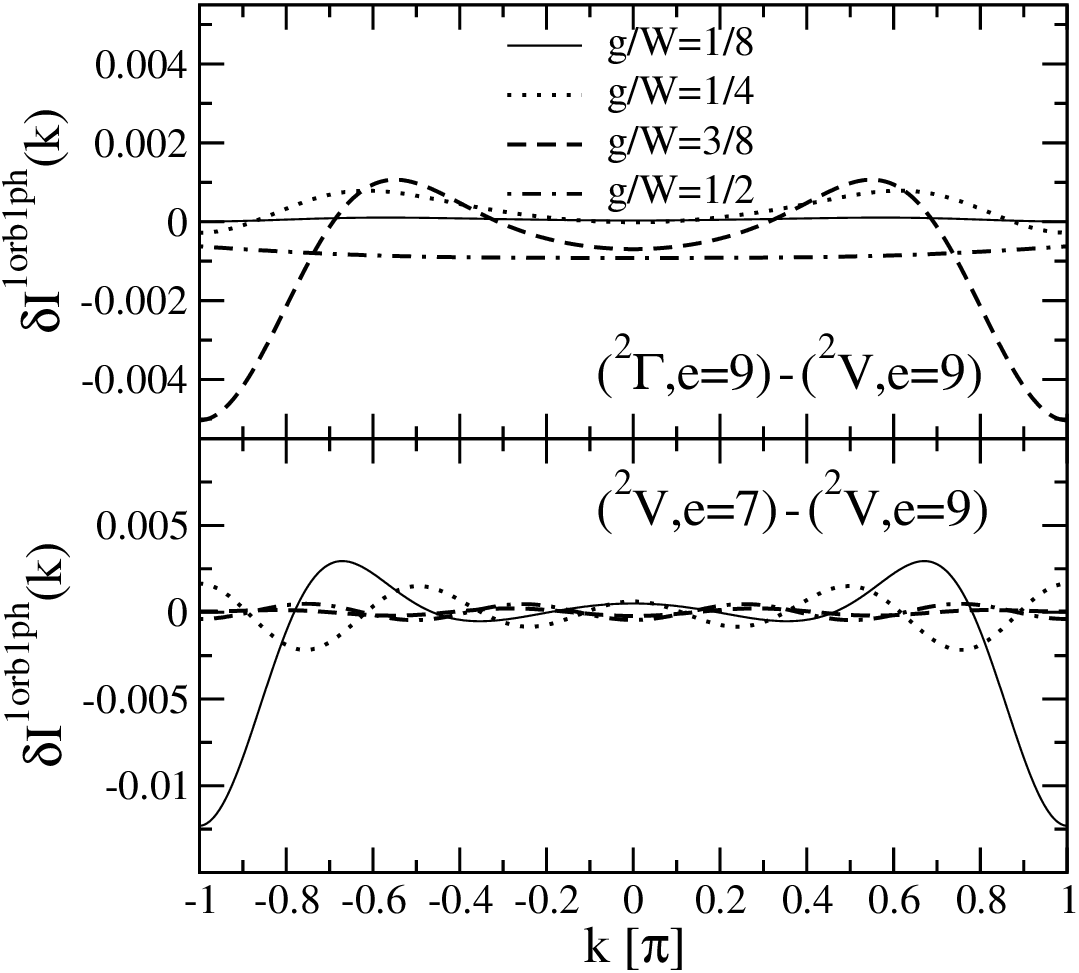}
    \end{center}
    \caption{Effect of the truncation on the $k$-resolved spectral weight of
      the orbiton-plus-one-phonon sideband for $\omega_{\rm ph}/W=1$.
      Upper panel: the difference $\delta I = I({^2\Gamma},e\!=\!9) \, - \,
      I({^2V},e\!=\!9)$ at constant extension $e=9$. Lower panel:
      $\delta I = I({^2V},e\!=\!7) \, - \, I({^2V},e\!=\!9)$ at constant
      structure $S={^2V}$.
    }
    \label{orbi-sw-k_1orb1ph_truncations}
\end{figure}

The commutators required for
the flow (\ref{eq:flow}) are computed using the standard bosonic 
algebra. The truncation of the proliferating terms in the flow equation
will be discussed below. We consider
 \begin{eqnarray}
  \label{eq:Ham_flow}
  H(\ell)&=&\sum_{i,n} J_n
 c^\dagger_{i+n}c^{\phantom{\dagger}}_i + \omega_{\rm ph}\sum_i b^\dagger_i
 b^{\phantom{\dagger}}_i
 \\
 \nonumber
&+& \sum_{i,n_1,n_2}  \left( \Gamma_{n_1,n_2} c^\dagger_{i+n_1}
 c^{\phantom{\dagger}}_{i+n_2}b^\dagger_i+{\rm h.c.}\right)
 \\  \nonumber
&+&\sum_{i,n_1,n_2,n_3}V_{n_1,n_2}^{n_3}c^\dagger_{i+n_1}
c^{\phantom{\dagger}}_{i+n_2}b^\dagger_{i+n_3}b^{\phantom{\dagger}}_i
 \\  \nonumber
&+&\sum_{i,n_j}\left(
 {}^2\Gamma_{n_1,n_2}^{n_3,n_4}c^\dagger_{i+n_1}
c^{\phantom{\dagger}}_{i+n_2}b^\dagger_{i+n_3}b^\dagger_{i+n_4}
b^{\phantom{\dagger}}_i +{\rm h.c.}\right)
 \\  \nonumber
&+&
 \sum_{i,n_j}{}^2V_{n_1,n_2}^{n_3,n_4,n_5}c^\dagger_{i+n_1}
c^{\phantom{\dagger}}_{i+n_2}b^\dagger_{i+n_3}b^\dagger_{i+n_4}
b^{\phantom{\dagger}}_{i+n_5}  b^{\phantom{\dagger}}_i \ .
\end{eqnarray}
The only finite amplitudes at $\ell$=0 are
$J_0(\ell\!  = \! 0)\! =\! \omega_{\rm orb}^{\rm ex}+
\omega_{\rm orb}^{\rm JT}$, $J_{1} (\ell \! = \! 0) \! = \! -J/4$,
and $\Gamma_{0,0} (\ell \! = \! 0) \! = \! 2g$.
The generator $\eta (\ell )$ is chosen to be
 \begin{eqnarray}
  \eta (\ell)&=& \sum_{i,n_1,n_2}  \left( \Gamma_{n_1,n_2} c^\dagger_{i+n_1}
 c^{\phantom{\dagger}}_{i+n_2}b^\dagger_i-{\rm h.c.}\right)
 \\ \nonumber
&+&\sum_{i,n_j}\left(
 {}^2\Gamma_{n_1,n_2}^{n_3,n_4}c^\dagger_{i+n_1}c^{\phantom{\dagger}}_{i+n_2}
b^\dagger_{i+n_3}b^\dagger_{i+n_4}b^{\phantom{\dagger}}_i
 -{\rm h.c.}\right)\ .
\end{eqnarray}
In the following we denote the considered truncations by $(S,e)$ with the
structure $S \in \{\Gamma, V, {^2\Gamma}, {^2V} \}$ and the spatial extension
$e$. The structure $S$ is defined by the maximum number of creation and
annihilation operators appearing in the flowing Hamiltonian, e.g.\ $S={^2V}$
for $H(\ell)$ given in Eq.~\ref{eq:Ham_flow}. The spatial extension $e$ is
defined for a given operator $c^\dagger_{i+n_1}c^{\phantom{\dagger}}_{i+n_2}
b^\dagger_{i+n_3}\ldots b^{\phantom{\dagger}}_i$ by the distance between the
leftmost and the rightmost local operator, i.e.\ for $e$=9
only the exchange amplitudes $J_n$ with $n \! \in \! \{ -4;-3;\ldots;3;4\}$ are
finite. We considered $e \! \in \! \{1,3,5,7,9\}$ for the numerical evaluation.
We have found that $e$=9 and $S={^2V}$ is sufficient to obtain numerically
stable, accurate results (see below, Figs.~\ref{orbi-disp_truncations},
\ref{orbi-sw-k_1orb_truncations}, and \ref{orbi-sw-k_1orb1ph_truncations}).

\section{Results}
\subsection{Orbiton Dispersion}
In Fig.\ \ref{orbi-disp}, the dispersion of the dressed orbitons is shown for
$\omega_{\rm ph} \!= \! W$ using the truncation $({^2V},e\!=\!9)$.
With increasing $g/W$ the effective band width
$W_{\rm eff}$ is strongly reduced (see inset). The orbiton becomes dynamically
dressed by a phonon cloud enhancing its effective mass. Already for $g/W$=1/2
we find $W_{\rm eff}/W$=0.35 in CUT. In contrast, the approximate perturbative
result $W_{\rm eff}/W$=0.56 obtained by SOPT clearly
underestimates the reduction of the band width. For $g/W$=3/4 we find
$W_{\rm eff}/W$=0.1 in CUT, a factor of 4 smaller than in SOPT. This indicates
that for $g \gtrapprox 0.75 W$ the excitation may be regarded as `local' for
practical purposes, i.e.\ rather as a vibron than as a propagating orbiton.

The dependence of the orbiton dispersion on the truncation level is depicted in
Fig.~\ref{orbi-disp_truncations}. The upper panel shows the difference
$\delta \omega(k) = \omega(^2\Gamma,e\!=\!9) \, - \, \omega(^2V,e\!=\!9)$,
illustrating the dependence on the structure $S$. Clearly, $\delta \omega(k)$
increases with $g$ but remains small ($\delta \omega / W_{\rm eff} < 1$\%)
for all considered values of $g$.
The truncation error resulting from the finite spatial extension $e$ of the
operators is even smaller, as shown in the lower panel of
Fig.~\ref{orbi-disp_truncations}. This reflects the fact that the physics of
the model becomes more local with increasing $g$, hence less extended operators
are sufficient to describe the orbiton dispersion. This is underlined by the
observation that the truncation error related to the extension $e$
is smaller for $g/W=1/2$ than for $g/W=3/8$.

\subsection{Spectral Properties}
\begin{figure}[tb]
    \begin{center}
    \includegraphics[width=0.9\columnwidth,clip=]{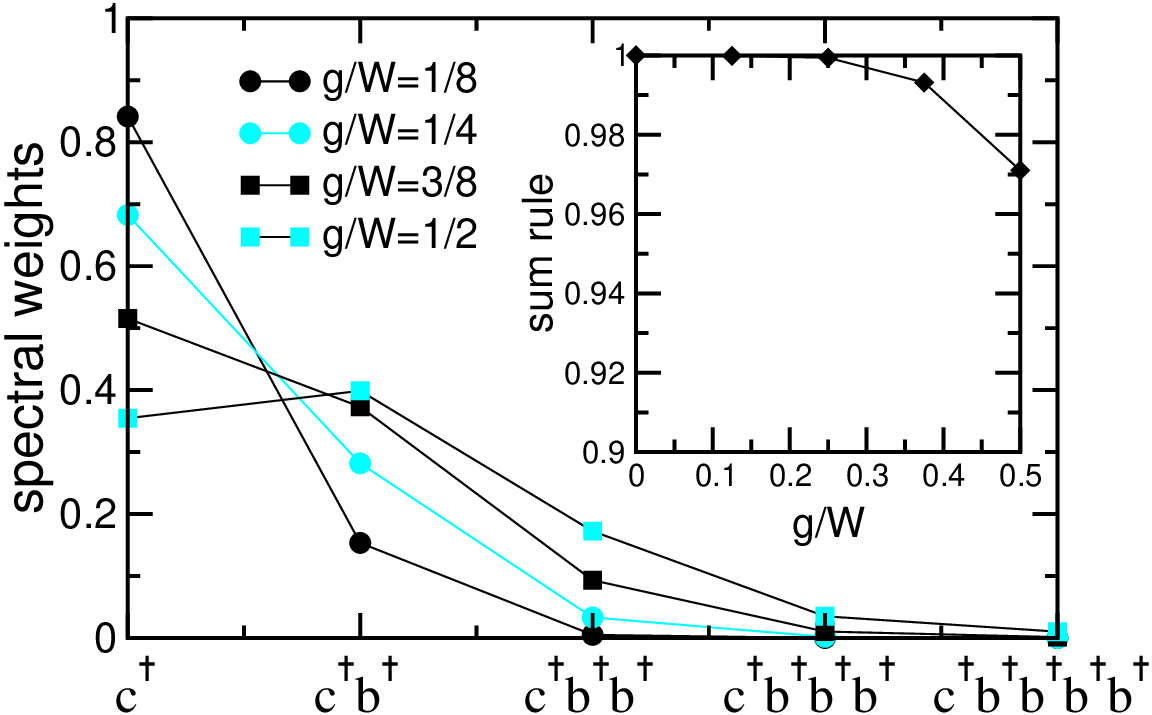}
    \end{center}
    \caption{(Color online) Spectral weights of the orbiton quasi-particle band and of the
    orbiton-plus-$n$-phonon continua ($n$=1-4) for $\omega_{\rm ph}/W=1$
    using the truncation $({^2V},e\!=\!9)$.
    Inset: The sum of the weights of the five considered contributions obeys
    the sum rule very well. Lines are guides to the eye only.
    }
    \label{orbi-sw}
\end{figure}

For the orbiton line shape, we transform the local orbiton creation operator
$\mathcal{O}^{\rm loc}=c_0^\dagger$ to an effective one by the same CUT as
applied to $H_0$.
We study all terms with one orbiton operator plus up to four phonon operators
\footnote{The number of phonon channels has been varied to verify that
a sufficiently large number of phonons is considered. This ensures that
the results do not depend much on this truncation. }
\begin{eqnarray}
\mathcal{O}(\ell)&=&\sum_r A_r c^\dagger_r
+ \sum_{r_1,r_2}\left( B_{r_1}^{r_2}c^\dagger_{r_1} b^\dagger_{r_2}
 +C_{r_1}^{r_2}c^\dagger_{r_1}b^{\phantom{\dagger}}_{r_2}\right)
 \\ \nonumber
 &+& \ldots
 \\ \nonumber
 &+& \sum_{r_j}\left( K_{r_1}^{r_2,r_3,r_4,r_5}c^\dagger_{r_1} b^\dagger_{r_2}
b^\dagger_{r_3} b^\dagger_{r_4} b^\dagger_{r_5} +\ldots \right) + {\rm h.c.}
\ .
\end{eqnarray}
Initially, the only finite value is $A_0 (\ell =0)=1$. Finally, at $T=0$, there
 are five types of contributions (omitting spatial indices):
$c^\dagger$,
$c^\dagger b^\dagger$,
$c^\dagger b^\dagger b^\dagger$,
$c^\dagger b^\dagger b^\dagger b^\dagger$,
and $c^\dagger b^\dagger b^\dagger b^\dagger b^\dagger$.
For each type we sum the moduli squared of all
coefficients, yielding the spectral weights given in Fig.\ \ref{orbi-sw}.

The sum of \emph{all} contributions has to equal unity. The inset of Fig.\
\ref{orbi-sw} shows that the sum of the five considered contributions is
very close to 1 for $g/W \leq 1/2$ so that higher-order terms are indeed not
important in this parameter range. For small values of $g/W$, the spectral
weight resides almost entirely in the fundamental orbiton band,
i.e., the eigen mode is a well-defined orbital wave.
With increasing $g/W$ spectral weight is transferred to the
orbiton-phonon continua, which reflects the change of the excitation character
from an orbital wave to a vibronic excitation with increasing $g$.
Already for $g/W$=1/2 the orbiton-plus-one-phonon band
dominates the spectrum. This agrees roughly with the result obtained
for the local limit of both orbitons and phonons \cite{brink01}.

\begin{figure}[tb]
    \begin{center}
     \includegraphics[width=0.9\columnwidth,clip=]{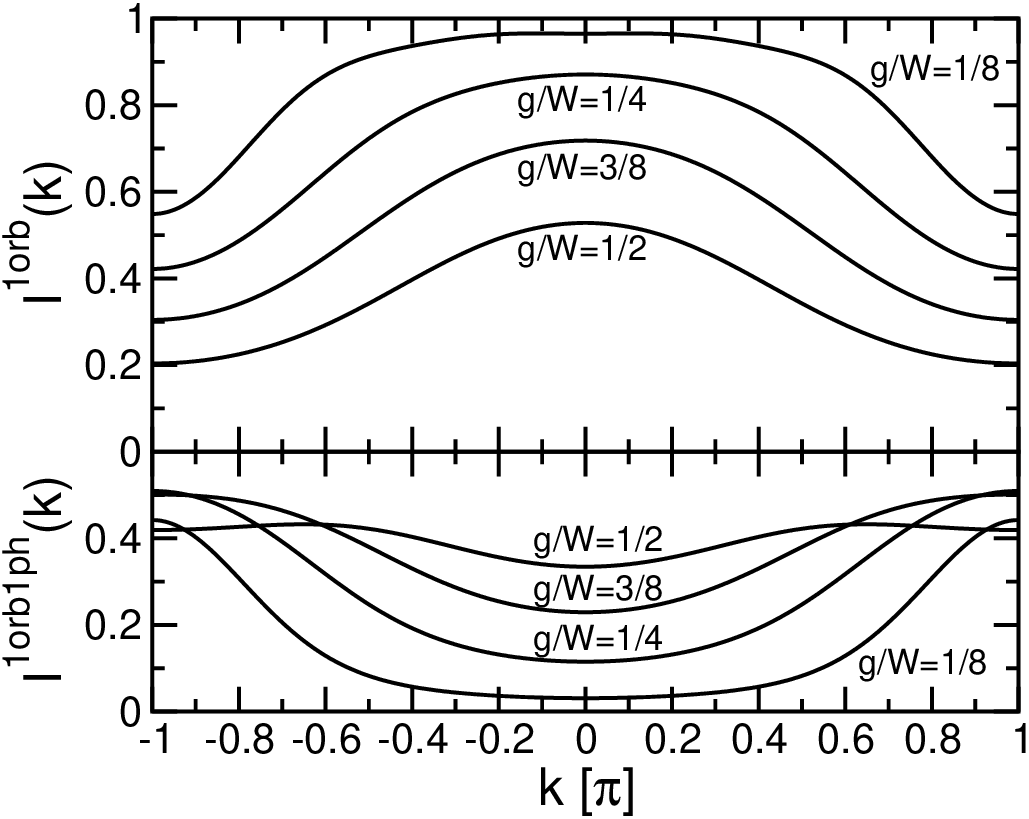}
    \end{center}
    \caption{The $k$-resolved spectral weight of the orbiton quasi-particle
      band (top) and of the orbiton-plus-one-phonon sideband (bottom) for
      $\omega_{\rm ph}\!=\!W$ using the truncation $({^2V},e\!=\!9)$.
    }
    \label{orbi-sw-k}
\end{figure}

\begin{figure}[tb]
    \begin{center}
     \includegraphics[width=0.9\columnwidth,clip=]{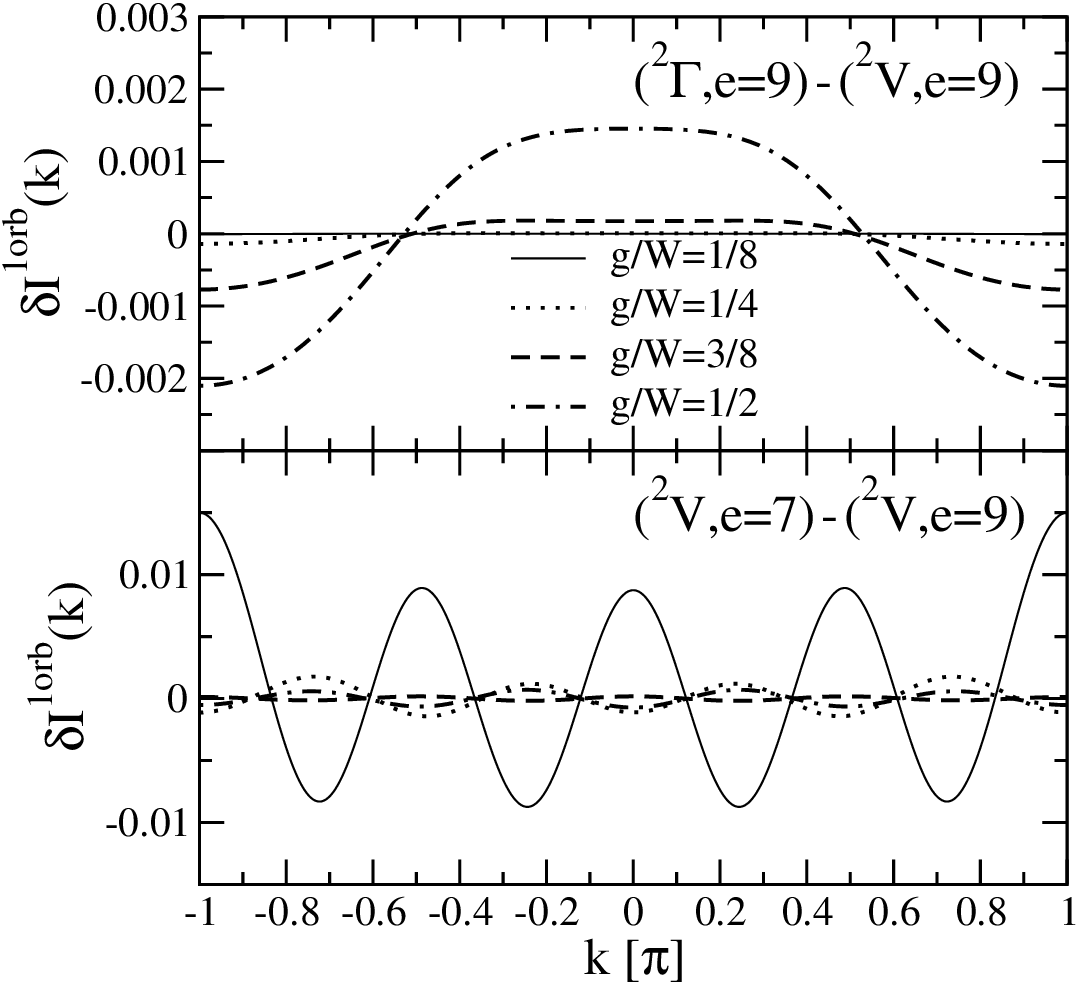}
    \end{center}
    \caption{Effect of the truncation on the $k$-resolved spectral weight of
      the orbiton quasi-particle band for $\omega_{\rm ph}/W=1$.
      Upper panel: the difference $\delta I = I({^2\Gamma},e\!=\!9) \, - \,
      I({^2V},e\!=\!9)$ at constant extension $e=9$. Lower panel:
      $\delta I = I({^2V},e\!=\!7) \, - \, I({^2V},e\!=\!9)$ at constant
      structure $S={^2V}$.
    }
    \label{orbi-sw-k_1orb_truncations}
\end{figure}

\begin{figure}[p!]
    \begin{center}
      \hspace*{-2.2mm}
      \includegraphics[width=0.86\columnwidth]{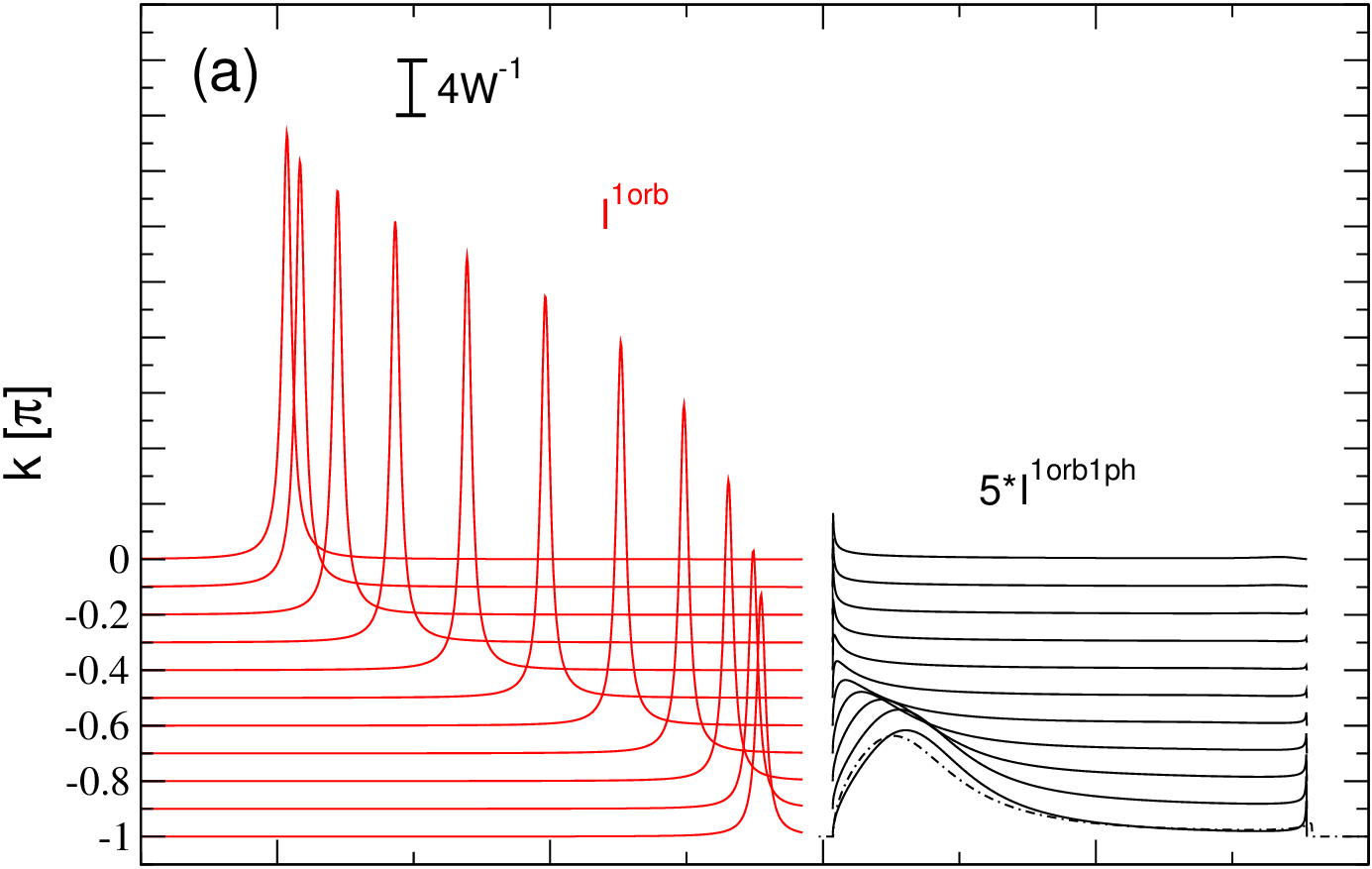}
      \hspace*{-2.2mm}
      \includegraphics[width=0.86\columnwidth]{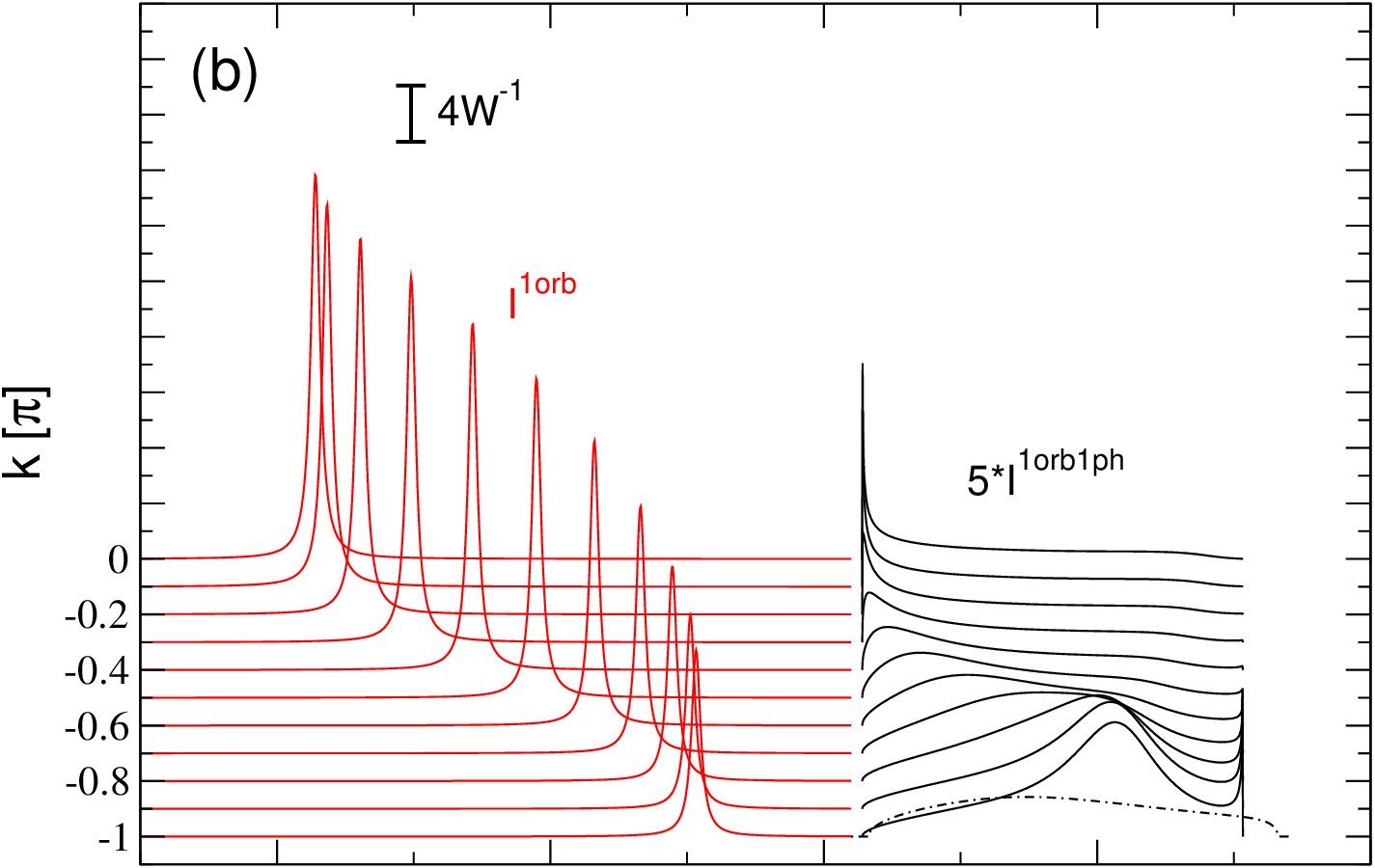}
      \hspace*{-2.2mm}
      \includegraphics[width=0.86\columnwidth]{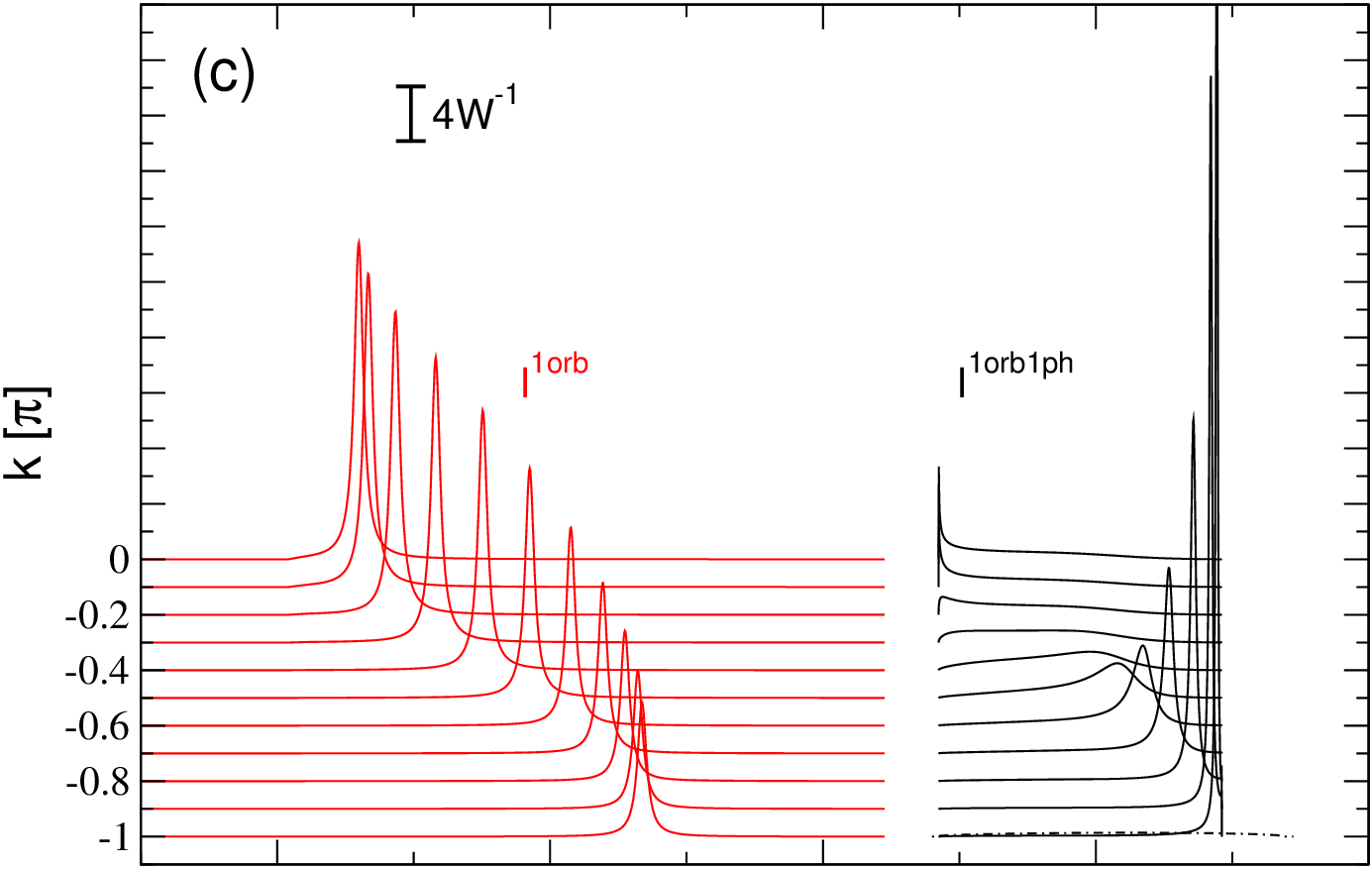}
      \includegraphics[width=0.875\columnwidth]{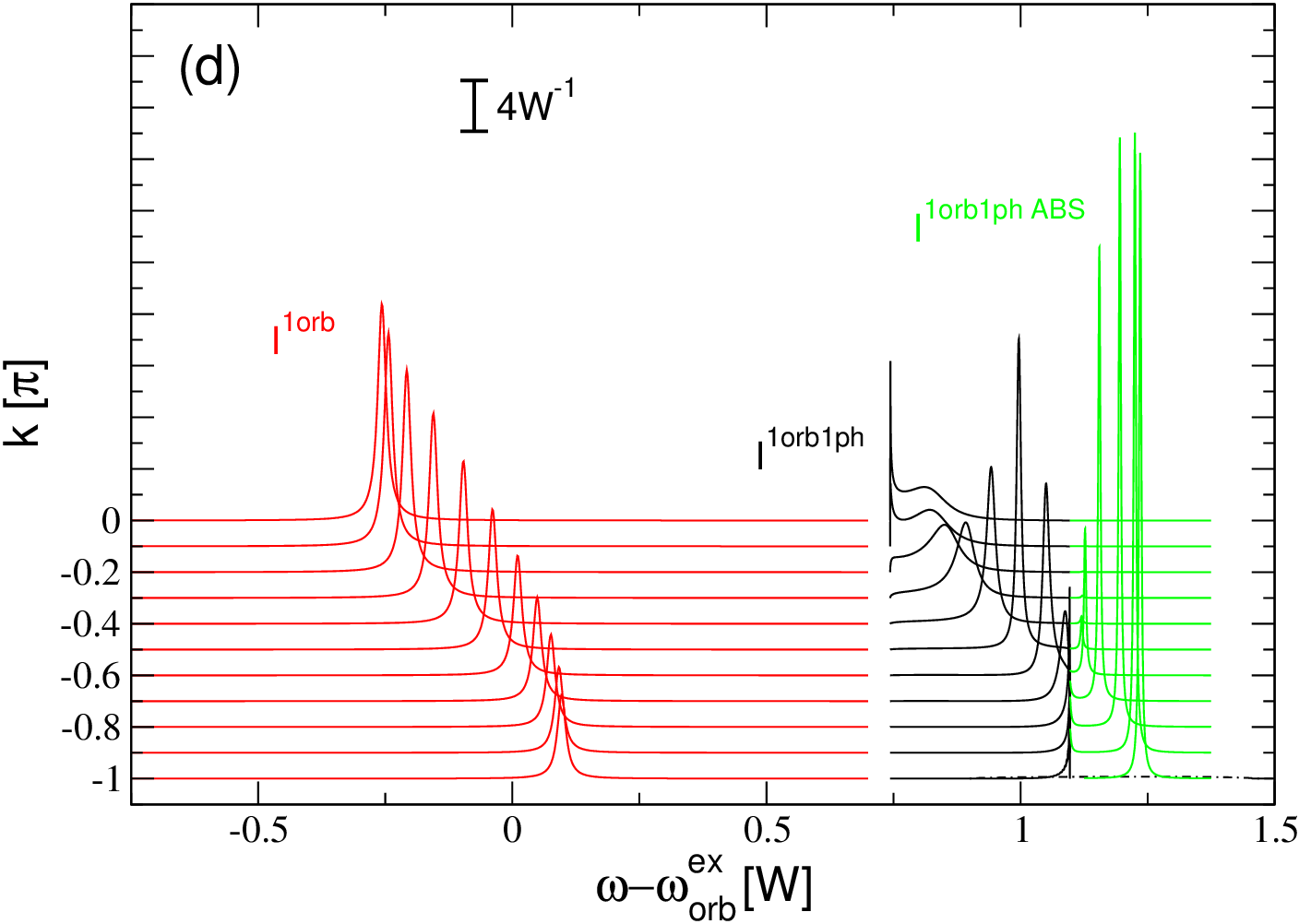}
    \end{center}
    \caption{(Color online) Spectral density of one orbiton (red) and
      the orbiton-plus-one-phonon continuum (black) for $\omega_{\rm ph}/W=1$
      using the truncation $({^2V},e\!=\!9)$.
      An orbiton-phonon anti-bound state (green) is
      shown where it has gained substantial weight.
      (a) $g/W$=1/8. (b) $g/W$=1/4. (c) $g/W$=3/8. (d) $g/W$=1/2.
      The $\delta$-functions are broadened by $\Gamma=0.01W$.
      The dashed lines depict the  orbiton-plus-one-phonon continuum
      in SOPT for $k=-\pi$.
    }
    \label{orbi-kont}
\end{figure}

Now we turn to momentum ($k$) and frequency ($\omega$) resolved spectral
properties. The $k$-resolved spectral weights of the dressed orbiton and of the
orbiton-plus-one-phonon continuum are plotted in Fig.\ 4. The dependence of
these two quantities on the truncation level is illustrated in Figs.\ 5 and 6.
Again, the differences between different truncations are small for all
considered values of $g$.
As mentioned above, the physics becomes more local with increasing $g$,
explaining the larger sensitivity on the spatial extension observed
for $g/W$=1/8.

The corresponding $k$-resolved spectral densities are shown in Fig.\ 7.
The transfer of spectral weight away from the orbiton band
is largest at the Brillouin zone boundary,
see top panel of Fig.\ 4, where the orbiton band is energetically close to
the continuum, see Fig.\ 7a. For small values of $g/W$, the
spectral weight `leaks' into the continuum where it appears as a broad hump.
For $g/W$=1/8, the results of SOPT and CUT for the line shape and the spectral
weight still agree well with each other, see Fig.\ 7a.
With increasing $g/W$, the effective orbiton band width $W_{\rm eff}$ is
reduced so that the separation between the orbiton band and the
orbiton-plus-one-phonon continuum increases. At the same time, a sharp
resonance appears within the continuum, displaying a clear dispersion.

For $g/W$=1/2, the orbiton-plus-one-phonon sector
dominates the spectrum, cf.\ Fig.\ 3. Its spectral weight is almost
independent of $k$, in contrast to the behavior observed for smaller values of
$g/W$, see bottom panel of Fig.\ 4.
The relative suppression of the orbiton-plus-one-phonon continuum close to the
Brillouin zone boundary reflects the transfer of spectral weight to the
orbiton-plus-{\em two}-phonon continuum.

\subsection{Anti-Bound State}
Interestingly, a very sharp dispersive feature is pushed out of the continuum
to higher energies. Because 
it is built from states with one orbiton and one phonon and because
it develops from the scattering states in the
continuum,
but lying at higher energies, this feature is identified
as an anti-bound state (ABS) of one orbiton and one phonon.

The appearance of an ABS is also found in SOPT,
though at different energies and with different weights (not shown).
\footnote{The occurrence of the ABS at \emph{arbitrarily} weak $g$ will be
specific to one dimension. But we expect an anti-bound state to occur in all
dimensions for sufficiently large orbiton-phonon interaction.}
It occurs for all momenta and arbitrarily weak couplings $g$ due to the 1D
van Hove singularities. In contrast, the sharp resonance seen within the
orbiton-phonon continuum at intermediate couplings is totally missed by the
SOPT which stays broad and featureless, see dashed lines in Fig.\ 7.

For $g/W$=1/2 the total dispersion  -- from the peak maxima in the
continuum at $k$=0 to the ABS at $k$=$-\pi$ -- is even {\em larger}
than the renormalized band width $W_{\rm eff}$ of the orbiton band.
This surprising result indicates that the dispersion of collective orbital
excitations may very well be observed experimentally even for intermediate
values of the phonon coupling $g$.

The dispersion of the anti-bound state and of the resonance in the
orbiton-phonon continuum comes as a surprise since we started from entirely
local phonons. How can an ABS of an orbiton and of an immobile phonon
propagate ?
Since in our model (\ref{eq:hamilton}) a single orbiton cannot influence
the macroscopic number of phonons we can exclude that some hopping amplitude
of the phonons is induced by $g$.
Hence we are led to the conclusion that there must be an important
momentum-dependent effective interaction between an orbiton and a phonon.
A pair of orbiton and phonon undergoes correlated hopping.
In the Hamiltonian $H_0$ in (\ref{eq:hamilton}) before the CUT, this hopping
can presumably be understood as a virtual process
where the orbiton-phonon pair
is  intermediately de-excited to a single orbiton which can hop.
Momentum-dependent matrix elements are certainly also present; but they alone
cannot explain the dispersion of the ABS.

In the literature on the Holstein model,
the appearance of sharp subbands which are
qualitatively similar to our results has been reported for $W$ significantly
larger than $\omega_{\rm ph}$ and large values of the coupling
constant.\cite{hohen03,fehsk06}
However, the identification of the anti-bound state by our CUT approach is
essential in order to understand why the dispersion in higher subbands is
larger than in the elementary orbiton/polaron band.

\section{Summary}
Our work is intended to provide information on the nature of orbital
excitations in the presence of substantial orbiton-phonon coupling
in principle. We do not aim at any particular compound. Therefore, we
have concentrated on a simplified model in one dimension
without hybridization or phonon dispersion. As indicated in the discussion
following Eq.~(\ref{eq:hamilton}), we expect that our findings apply
qualitatively also to more specific, extended models.
In two or three dimensions, the line shape at the band edges will
change and the anti-bound state will form only beyond a certain
threshold of the orbiton-phonon interaction. The presence of hybridization
will imply that the orbiton line shape occurs as a weak feature
also in experimental probes coupling directly to the distortions
and vice versa. Finally, a finite phonon dispersion will contribute
to the mobility of the collective states. In case continua start to
overlap, effects of finite life times due to decay will be observable.
These points summarize what we expect for the modifications of
our findings in real systems.

We investigated the gradual transition from a propagating orbital
wave to a `local' vibron with increasing phonon coupling $g$.
We found already for intermediate couplings a substantial
reduction of the orbiton band width.
The orbiton-phonon continuum is not a featureless hump
as suggested by second order perturbation theory (Born approximation),
but displays a relatively sharp, dispersive resonance
and an anti-bound state. For $g/W$=1/2, the orbiton-plus-one-phonon
sector carries more spectral weight than the orbiton itself, but it
also shows the larger dispersion. Thus the collective character is seen rather
in the orbiton-phonon sector than in the fundamental orbiton sector.
This  may turn out as a major advantage for experiments, since phonons and
orbiton-plus-phonon features appear in separate frequency ranges, facilitating
the correct assignment.
Our results demonstrate that signatures of collective orbital excitations are
not limited to compounds in which the phonon coupling $g$ is very small.

We acknowledge fruitful discussion with H.~Fehske and G.~Khalliulin and the
financial support of the DFG in SFB 608 in which this project has been started.


\end{document}